%% file: SPR_Final.tex
\documentclass{elsart}

\journal{Physica A}

\usepackage[numbers]{natbib}
\usepackage{subeqn}
\usepackage{graphicx}
\usepackage{amssymb}

\begin{document}
\date{3 May 2006}
\begin{frontmatter}

\title{Determinism, Noise, and Spurious Estimations in a Generalised Model of Population Growth}

\author{Harold P. de Vladar\corauthref{cor1}}
\ead{H.Vladar@biol.rug.nl}
\author{Ido Pen}
\ead{I.R.Pen@rug.nl}

\corauth[cor1]{Corresponding Author}
\address{Theoretical Biology Group, CEES\\
University of Groningen\\
Kerklaan 30. 9751NN Haren. The Netherlands.}

\begin{abstract}
We study a generalised model of population growth in which the state variable is population growth rate instead of population size. Stochastic parametric perturbations, modelling phenotypic variability, lead to a Langevin system with two sources of multiplicative noise. The stationary probability distributions have two characteristic power-law scales. Numerical simulations show that noise suppresses the explosion of the growth rate which occurs in the deterministic counterpart. Instead, in different parameter regimes populations will grow with ``anomalous'' stochastic rates and (i) stabilise at ``random carrying capacities'', or (ii) go extinct in random times. Using logistic fits to reconstruct the simulated data, we find that even highly significant estimations do not recover or reflect information about the deterministic part of the process. Therefore, the logistic interpretation is not biologically meaningful. These results have implications for distinct model-aided calculations in biological situations because these kinds of estimations could lead to spurious conclusions.
\end{abstract}


\end{frontmatter}
\bibliographystyle{unsrtnat}

Population dynamics are frequently modelled with simple equations that mimic some aspects of replicating biological entities, such as division (in cells), fission (in modular organisms) or reproduction (in eukaryotes), competition, and population-size limiting (saturation). These and other properties are represented by various models. Frequently the validity of these models is a matter of statistical goodness of fit with a specific data set. However, these biological properties are not entirely of intrinsic nature to the individuals, or to the populations themselves, but rather emerging ecological properties, i.e. the interaction between ``individuals'' and ``environment''. The models of population growth simplify (whenever it is possible) the potential complexity of a detailed ecological description into simple equations.

Recently we showed that a variety of biological growth models can be unified using a phase-space decomposition using two dynamical variables, population size  $x$ and growth rate $r$ \citep{PdeVladar06a} (analogous to a particle' s position and momentum, respectively):
\begin{subequations}
\label{syst:global} 
\begin{equation}
\dot{x} = xr~,
\end{equation}
\begin{equation}
\label{eq:rate}
\dot{r} = r(\theta r -\rho).
\end{equation}
\end{subequations}

The constant $\rho$ is the Malthusian parameter, and $\theta$ is the intraspecific interaction coefficient. By varying these two parameters it is possible to reproduce exactly a wide family of growth laws including exponentials, logistics \citep{Ayala73,Sibly05}, Gompertzian \citep{Kozusko03}, Potential \citep{Roff86,Szathmary87}, as well as allometric growth laws like Von Bertalanffy's \citep{VonBertalanffy66,Roff86}  and West's \citep{West02} equations, among others \citep{PdeVladar06a}.

Although the deterministic behaviours can be associated with distinct biological scenarios \citep{PdeVladar06a}, populations are often influenced by some source of noise. Biologically, random ``forces'' are often related to environmental and demographic fluctuations, as well as to intrinsic complex effects like genetic variability, mating, and segregation. These environmental and genetic factors can be thought to be the determinants of the parameters that describe the growth of a population. This conjecture means that they are (complicated) measures of phenotypic expression. Thus to a first approximation we can model the effect of phenotypic variation as noise over these parameters.

\newcommand{\etat}{\eta_{\theta~(t)}}
\newcommand{\etar}{\eta_{\rho~(t)}}
We can consider two sources of noise $\etar$ and $\etat$ affecting respectively the parameters $\theta$ and $\rho$. The rate equation (\ref{eq:rate}) linearly perturbed with \(\rho \rightarrow \rho + \etar\) and \hbox{\(\theta \rightarrow \theta+\etat\)} results in
\begin{equation}
\label{eq:stochrate}
\dot{r} = r (\theta r-\rho) + r^2 \etat + r \etar~,
\end{equation}
where the \(\eta_{i~(t)}\) have the usual properties of white noise:
\begin{equation}
\begin{array}{c}
    \langle \eta_{i~(t)} \rangle = 0 \\
    \langle \eta_{i~(t)} \eta_{i~(s)}\rangle = \delta(t-s) \epsilon^2_i\\
    \langle \etat \etar \rangle = \gamma\\
\end{array}
\end{equation}
with \(\langle \ldots\rangle\) denoting expectations, and \(i = \rho,\theta\). Here, \(\epsilon_i\) are the intensities of the noise sources, $\gamma$ is the correlation between the two noise sources, and \(\delta(t)\) is Dirac's delta function.

The resulting system is a Langevin equation where \(\rho\) is the drift term
and \(\theta r^2\) can be thought as the force of an external field
\cite{Gardiner}. Multiplicative noise, often represents fluctuating
barriers or processes of anomalous diffusion (i.e. diffusion where the probability of long steps is higher then in the normal case) \citep{Biro05,Fa03,Kaniadakis03,Hanggi94,Fleming93}. Also, multiplicative noise is a process that retains memory (i.e. is non-Markovian), and has been investigated in the context of population growth and extinctions \citep{Wichmann05,Ai03,Halley99}.
 
 The stochastic differential equation (SDE, Eq. \ref{eq:stochrate}) remains uncoupled from the size $x$. This gives an operational advantage since the analyses of the SDE can be made in terms of $r$ as a 1-dimensional system that is relatively simple to handle.

To study the effects of multiplicative noise, and make precise the meaning of ``anomalous growth'' in populations (in analogy to anomalous diffusion), first consider the probability distribution for the rates. In the It\^o interpretation of noise the probability is given by the related Fokker Planck equation (FPE):
\newcommand{\prt}{\mathcal{P}_{(r,t)}}
\newcommand{\pr}{\mathcal{P}_{(r)}}
\newcommand{\epste}{\epsilon_\theta^2}
\newcommand{\epsro}{\epsilon_\rho^2}
\newcommand{\roo}{\rho_0}
\newcommand{\teo}{\theta_0}
\begin{equation}
\label{GFPE}
\partial_t \prt = - \partial_r \left[ r(\theta r-\rho) \prt \right] + \frac{1}{2} \partial_{rr} \left[(\epsro-2 \epsilon_\rho \epsilon_\theta \gamma r+\epste r^2)r^2 \prt\right]~.
\end{equation}
Setting the time derivative equal to zero makes it possible to calculate the potential solution of the FPE on the stationary regime $P(r)$ (i.e. equilibrium solution), which gives
\begin{equation}
\begin{array}{rcc}
\label{SFPE}
P(r) &:=& \mathcal{N} \left(r^{-2}\right)^{\roo +1}\left( \epsilon_0^2 - \epsilon_0 \gamma r + r^2 \right)^{\roo - 1} \times\\
&\times&\exp\left[2(\epsilon_c \teo- \gamma_c \roo ) \tan^{-1} \left(\epsilon_c r -\gamma_c \right)\right]~,
\end{array}
\end{equation}
where \(\mathcal{N}\) is the integration constant, and
\begin{displaymath}
\begin{array}{lll}
\roo = \rho/\epsro~, & \teo = \theta/\epste~, &\epsilon_0 = \epsilon_\rho/\epsilon_\theta~,\\
\epsilon_c = \epsilon_0^{-1} (1-\gamma^2)^{-1/2}~, & \gamma_c = \gamma (1-\gamma^2)^{-1/2}.&\\
\end{array}
\end{displaymath}

Fig. \ref{fig:SFPEsols} shows that there are three distinct kinds of stationary distributions for the rates. The first thing that we note, is that the correlation $\gamma$ modulates the transition from one distribution to another. Thus, for simplicity for the further analyses we proceed setting $\gamma=0$. 

The first distribution (Fig. \ref{fig:SFPEsols}A) is monotonous decreasing, and the other two (Figs. \ref{fig:SFPEsols}B-C) have an analytic maximum at \[r^* = \frac{\teo}{4} + \sqrt{\left(\frac{\teo}{4}\right)^2-\epsilon_0^2 (\roo+1)} ~.\] The condition for having an analytic maximum is
\begin{equation}
\label{eq:PertSpace}
\frac{\teo^2}{8} \geq \epsilon_0^2 (\roo + 1)~.
\end{equation}

Fig. \ref{fig:paramspace} outlines the regions where the inequality (\ref{eq:PertSpace}) holds. The parabolic curve (i.e. the boundary given by the equality in Eq. \ref{eq:PertSpace}) divides the parameter space into three regions with different stationary regimes. The first, when \( 0 < \teo^2/8 < \epsilon_0^2 (\roo + 1)\), corresponds to the space under the parabola (Fig. \ref{fig:paramspace}); in this case the probability density is accumulated at $r=0$. The second region, characterised by \( 0 < \epsilon_0^2 (\roo + 1) \leq \teo^2/8\), corresponds to the space over or under the parabola region (Fig. \ref{fig:paramspace}). The third region, defined by \( \epsilon_0^2 (\roo + 1) < 0 \leq \teo^2/8\), is the space on the left of the parabola. In the two last regions, the probability mass of rates is distributed along the axis, indicating that the growth rate can be persistent (i.e. non-zero). In the following of the paper, we will show that each of these regions have distinct qualitative solutions in which the deterministic nature of the process is ``forgotten'', but the resulting dynamics of the population size look like exponential or logistic dynamics. We will demonstrate however, that these two forms are entirely product of noise, hence fitting these models to the realizations -although statistically significant- are spurious.

For particular cases of Eq. (\ref{eq:stochrate}) the stationary distribution has been calculated before. When $\theta$ and its noise $\eta_\theta$ term are absent, the equation recovers the geometric Brownian motion \citep{Oksendal}, whose stationary distributions were shown to have power-law tails \citep{Biro05}. In this representation, the growth corresponds to a Gompertzian growth. A power-law-tailed distribution is also found for the stationary distribution of an equation where \(\rho \neq 0\) but which is not perturbed \citep{Gora05}. Also, a logistic case \(\theta=1\) was analysed by \citet{Morita86} using perturbation techniques for a time dependent solution. In log-log scale the distribution (\ref{SFPE}) is kinked near \(r_c = \epsilon_0 \exp\left(\pi \theta_0/2 \epsilon_0 (1-\rho_0)\right)\), with a right-tail decreasing in a power law fashion \(\log P(r) \sim -4 \log r\) (insets in Fig. \ref{fig:SFPEsols}). This is a result that can be derived directly from the particular case studied by \citet{Gora05}, because the right tail of the distribution is independent of the parameters. Moreover, there is also a power law behaviour for small values of $r$ which is given by \(\log P(r) \sim -2(\roo+1) \log r\). These power law tails lead to Tsallis statistics \citep{Biro05,Anteneodo03}. Some relationships between exponents have been derived for a system related (but not equivalent) to ours \citep{Genovese99}. 

The fast decrease of the right tail has an important consequence, which is the boundedness of the process. In other words, it means that the fluctuations remain finite. The simulations of Fig. \ref{fig:RealizationR}A show that when the probability density is accumulated at $r=0$ the rates will stochastically reach zero and stay there forever. Whenever this happens, the population freezes at its -random- current size. 

\citet{Sibly05} performed an analysis where they fitted more than a thousand population time series to the $\theta$-logistic model. Their analysis was based on the size-dependent per capita growth rate $r(x)$. The examples they presented, show comparable patterns to the realizations obtained from our model (Fig. \ref{fig:RealizationR}A). However, as we can see in this figure, the stochastic trajectories are notcentredd on the deterministic trajectories, as it is common for multiplicative perturbations. Therefore, the interpretation of the estimations in Ref. \citep{Sibly05} differ from the deterministic path, at least in the light of our model. We will return to this discussion later in the article.

Other simulations, using a processes having stationary distribution with maxima, are shown in figure \ref{fig:RealizationR}B-C. In this cases the rate does not explode in the time-window, even when these realizations (the deterministic and the stochastic) have the same initial conditions which would lead to explosions in the absence of perturbations. 

Recently, \citet{Mao02} demonstrated that the deterministic explosions of ``positive'' logistic equations (e.g. of the form \(\dot{x} = ax(1+bx)\)) can be controlled with certain types of multiplicative noise sources. When these fluctuations are present, populations will not diverge in finite time, although their purely deterministic analogue does. The rate-representation introduced in this paper is also of quadratic form, thus the results of \citet{Mao02} apply to Eq. (\ref{eq:stochrate}). However, the biological interpretations change, because explosions are suppressed in the rate rather than in population size.

As indicated by the distribution of the rates, probability is accumulated near the maximum, thus the rates will be non-vanishing, jumping from very slow to high (but finite) values, making the size of the population increase in bursts, reconstructing a devil's staircase pattern (a staircase where all the steps are of different size and height). Also, because the rate never reaches zero, the population grows unlimited.

The same distributions of Fig. \ref{fig:SFPEsols} appear for negative rates. The course of the population is the opposite, i.e. decreasing, although the distribution is the same (in absolute value): (i)if the distributions have an analytic maximum, the rates will remain finite and fluctuating, meaning that population will decrease erratically but monotonously and therefore populations will become extinct in random times; (ii) if the distributions do not have an analytic maximum, then $r$ reaches zero stochastically (Fig. \ref{fig:RealizationR}A), and then the populations will stabilise, again at a ``random carrying capacity''.

The distribution (\ref{SFPE}) is not normalisable whenever \(\roo+1>0\), because it diverges when \(r\rightarrow 0\). This means that $r=0$ is an exit barrier, and hence once the rate reaches zero it will stay there. This limit is the same if taken from the left, thus the rate cannot either jump to a negative value once it reaches zero. For instance the rates maintain their sign or become null, but never change sign. The meaning is that when the rates are stationary, an initially growing populations will continue to grow, or at most, cease growing but they will not suddenly shrink. Therefore, the converse is also true: populations that started shrinking, will not suddenly change its course and grow. They will continue to shrink until extinction, or reach a stable value.

The dynamics results in distinct realizations that can give drastically different solutions, when compared, for example, to the deterministic solution.  Fig. \ref{fig:RealizationX}A show that the equilibrium value of the populations can be very different from the deterministic carrying capacity. Thus the observed equilibrium value of the populations is no longer determined by the initial conditions, as in the deterministic case \citep{PdeVladar06a}. Actually, the carrying capacity is now a random variable. For example, in Fig. \ref{fig:RealizationX}A the size equation is solved for several realizations of the process (\ref{eq:stochrate}). For the naive eye, the distinct realizations could be seen as distinct ``noisy logistics'' with different carrying capacities. Comparing the realizations to a logistic equation gives highly significant fits, even when the data come from a common process having the same values of the parameters.

In order to determine if we can recover deterministic information of the processes, we performed simulations of 250 randomly selected values of \(\rho,\theta\) and for each we performed 30 realizations. To every growth curve we least-squares-fitted a logistic model, and calculated its parameters $\hat{\rho}$ and $\hat{x}_\infty$. Fig. \ref{fig:RealizationX}B shows a scatter plot of the estimated vs. the deterministic values of the Malthusian parameter, showing a poor relationship. Other correlations are presented in Table 1. These results show that the reconstructions are totally spurious since they do not reflect any information of the generating process. But because the dynamics resemble a logistic realization, accepting a null hypothesis that the biological phenomena determining growth are of logistic nature is true, statistically speaking. However, our calculations show that stochastic processes can account for the same qualitative and quantitative description. Therefore, simple analysis like least squares fits are not enough to confirm the logistic hypothesis.

There is an analogous effect for the case when the rates are persistent. Once the stochastic rates are in stationarity, the resulting population dynamics resembles exponential growth. In the deterministic exponential growth, the initial condition of the rate determines the growth parameter \cite{PdeVladar06a}. However, under our scheme, the growth process is Markovian, thus the initial conditions of $(x,r)$ do not affect the stationary distribution. As a consequence, the expected or averaged rates are spurious estimators of an exponential dynamic (Fig. \ref{fig:RealizationX}C). A similar problem was described by Renshaw \citep{Renshaw91} when demographic stochasticity is present in an exponentially growing population.

At this point it is necessary to make a distinction between the outcomes of noise sources coming from demographic or phenotypic stochasticity. The first has been studied and experimentally supported \citep{Renshaw91,Lande03}. This kind of stochasticity is such that randomness affects the population through events of accidental mortality or occasional migrations (and is analogous to energy input coming from a heat bath.) In these cases, the populations would fluctuate, for example, close to carrying capacities, and thus information for the deterministic part of the dynamics can be extracted by averaging. The second type, i.e. parameter stochasticity, is more related to fluctuations in phenotypes, which results from the ``superposition'' of genetic and environmental processes. However, from the perspective of our model, where carrying capacities are not an intrinsic property of the environment, this averaging might not make biological sense. As we said, an average of the stochastic trajectory does not recover the deterministic path, like in Figs. \ref{fig:RealizationX}A, \ref{fig:RealizationX}A. Of course, populations might still be subject to demographic stochasticity, and therefore show fluctuations around a stable size. In this case, we would be presented with an additional noise source $\eta_M$, more related to the measuring techniques, perturbing the size equation as:
\begin{equation}
\dot x = xr + \eta_{M}~,
\end{equation}
 that gives the fluctuating pattern over the stable size. (This is a problem known as \textit{filtering}: when the measuring procedure has additional noise sources, not taking them into account in the estimations, may bias the interpretation of the underlying process \citep{Oksendal}.) Considering this source of fluctuations is more related to time series estimation than to the biological aspects of our model. Therefore we shall not discuss it in further detail, and we defere the reader to Ref. \citep{Siefert03} for a method to overcome these kind of problems.
  
 To summarise, we have presented an analysis of a novel population growth model that is based in fluctuations in the per-capita growth rate, rather than in the growth variable. The result, is that the rate always remains finite, either because rate explosions are suppressed, or because rate is damped to zero stochastically. As a consequence, and depending on the relationship between the deterministic parameters and noise, the model reproduce patterns that resemble exponential and logistic (sigmoid) growth.  It is important to notice that these behaviours are irrespective on how the deterministic population would grow. These forms are determined by the fluctuations and not from the biological processes of birth and death, at least not in the conventional interpretationn and description. When \( \teo^2/8 \geq \epsilon_0^2 (\roo + 1)\) then the resulting population grows anomalously, but with bounded fluctuations, and resembles an exponential growth. When \( \teo^2/8 < \epsilon_0^2 (\roo + 1)\) then the populations grow toward saturation. However, this result challenges the idea of a carrying capacity, that is supposed to describe self regulatory processes and an intrinsic property of the environment. Here, it is an emerging property from the fluctuations. In both cases, and more critically in the second (the logistic), statistical fits to the realizations are highly significant. But since the effects of randomness override the deterministic forces of the system, making these statistical estimations becomes unreliable. In the context of our formulation, the question about rate estimations looses its sense, because forecasting using the classic deterministic models proves useless. Thus fluctuation analysis might prove more informative about the stochastic driving forces. In this way, estimations and forecasting can give other statistical solutions to classical and new problems, using our different perspective, that is, when populations are subject to phenotypic stochastic variability.

\include{SPRtables}
\include{SPRgraphicscaptions}
\include{SPRgraphics}

\end{document}

%% file: SPRtables.tex
\begin{table}[h]
\caption{The correlation coefficient between the parameters and estimators for the logistic estimations are very low, indicating that the fits have little predictive power with respect to the deterministic part of the process, even thought they could be considered virtually perfect.}
\begin{tabular}{l|ccc}
	 	& $\rho$   &  $\theta$ & $\hat{\rho}$	\\\hline\hline
$\hat{\rho}$ 	& 0.532    &  0.301    &		\\
$\hat{x}_\infty$& -0.0712  & -0.0290   &  -0.0119	\\
\end{tabular}
\label{table:correlations}
\end{table}

%% file: SPRgraphicscaptions.tex
Figure Captions.

\begin{figure}[h]
\caption{Potential solutions for the Stationary Fokker Planck equation. (A) When  \( 0 < \teo^2/8 < \epsilon_0^2 (\roo + 1)\), the distribution shows an exit barrier at $r =0$; in this example \(\rho = 1,~\epsilon_\rho =0.5,~\theta = 1,~\epsilon_\theta=0.1\). (B) When \( 0<\epsilon_0^2 (\roo + 1) \leq \teo^2/8\), besides the exit barrier, the distribution also shows an analytic maximum; in this example \(\rho =1,~\epsilon_\rho=0.4,~\theta=4,~\epsilon_\theta=1.\) (C) When  \( \epsilon_0^2 (\roo + 1) < 0 \leq \teo^2/8\), the exit barrier disappearss, and the distribution is unimodal with an analytic maximum; the parameters are \(\rho=-3,~\epsilon_\rho=0.5,~\theta=2,~\epsilon_\theta=1.5\). The insets plot the distributions in log-log scale, showing that there are two characteristic scales. The left tail scales with an exponent of \(-2(1+\rho_0)\) whereas the right tail scales with an exponent of $-4$. The inflection points are close to the maximum.}
\label{fig:SFPEsols}
\end{figure}

\begin{figure}[h]
\caption{The parameter space $(\rho,\theta)$ consists of four quadrants,
  corresponding to their sign combinations. The dotted lines indicate distinct deterministic growth functions known in the literature:  Potential (P) $\rho = 0$; logistic (L) $\rho > 0, \theta=1$; Gompertzian (G) $\rho > 0, \theta = 0$; West (W) \(\rho > 0, \theta=-1/4\); von Bertalanffy (VB) \(\rho > 0, \theta = -1/3\); Exponential (E) \(\rho =0, \theta =0\). The solid curve represents the noise-transition points between the three distinct regimes of the distributions of the rate: (a) inside the parabolic region  \( 0 < \teo^2/8 < \epsilon_0^2 (\roo + 1)\); (b) above or below the parabolic region \( 0<\epsilon_0^2 (\roo + 1) \leq \teo^2/8\); and (c) at the left of the parabolic region \( \epsilon_0^2 (\roo + 1) < 0 \leq \teo^2/8\).}
\label{fig:paramspace}
\end{figure}

\begin{figure}[h]
\caption{Realization of noisy rate dynamics. The dotted lines show deterministic solutions, while the continuous bold lines show the stochastic realizations. It can be seen that when the deterministic rates decrease to zero, the stochastic dynamics will also decrease to zero. Also, when deterministic rates explode the stochastic dynamics remain finite. These realizations correspond respectivelyy to the regimes and parameters of Fig. \ref{fig:SFPEsols}}
\label{fig:RealizationR}
\end{figure}

\begin{figure}[h]
\caption{(A) Integrations for population size for distinct realizations of the same  process when \(\teo^2/8 < \epsilon_0^2 (\roo + 1)\) using the parameters \(\rho=1,~\epsilon_\rho=0.5,~\theta=1,\epsilon_\theta=0.5\). The bold line shows the deterministic dynamics and the thin lines are realization for the population size. The dotted lines are logistic estimations. In this cases, the estimated  Malthusian parameters range between \(\hat{\rho} \in  (0.8,1.5)\). Carrying capacities range from \(\hat{x}_\infty \in (1,10^3)\), and the deterministic carrying capacity is \(x_\infty = 10.0\). All of the estimations have a regression coefficient \(R^2 > 0.995\) with p-values less than $10^{-3}$. (B) Correlation between the estimated and generating Malthusian parameters, from 7500 simulation spanningg 250 distinct pairs of uniformly distributed values of \(\rho \in[0,2],~\theta = \pm  \epsilon_0 \sqrt{2(\rho_0+1)}\), using \(\epsilon_{\rho} = 1.0,~\epsilon_{\theta} = 0.1\). The radii of the circles are $10^{-3} \log(SE)$ (\textit{SE} = standard error). The continuous line is the linear trend, which gives \(\hat{\rho} = 0.389591 + 0.371044\rho\) with \(R^2 = 0.2873\) (p\(<2.2~10^{-16}\)). The dotted line is a linear trend weighted with the inverse of the standard error of each estimation: \(\hat{\rho} = 0.91300 + 0.22528\rho\) with \(R^2 = 0.3393\) (p\(<2.2~10^{-16}\)). Comparing these two estimations we see that even in the best case (the weighted regression) the predictive power is poor. (C) Integration for population sizes for distinct realizations of the process when \(\teo^2/8 \geq \epsilon_0^2 (\roo + 1)\) using the parameters \(\rho=-3,~\epsilon_\rho=0.3,~\theta=2,\epsilon_\theta=1.5\)). The bold line represents the deterministic dynamics, and the thin lines the realizations for population size. The dotted lines are estimations for the exponential growth. The estimated values for the exponential growth parameter are in the range of \((21.8,26.0)\). All the estimations have a regression coefficient $R^2>0.998$ with p-values less than $10^{-16}$. The graph is in semi-log scale.}
\label{fig:RealizationX}
\end{figure}

%% file: SPRgraphics.tex

\newpage

\includegraphics[scale=0.5,angle=0]{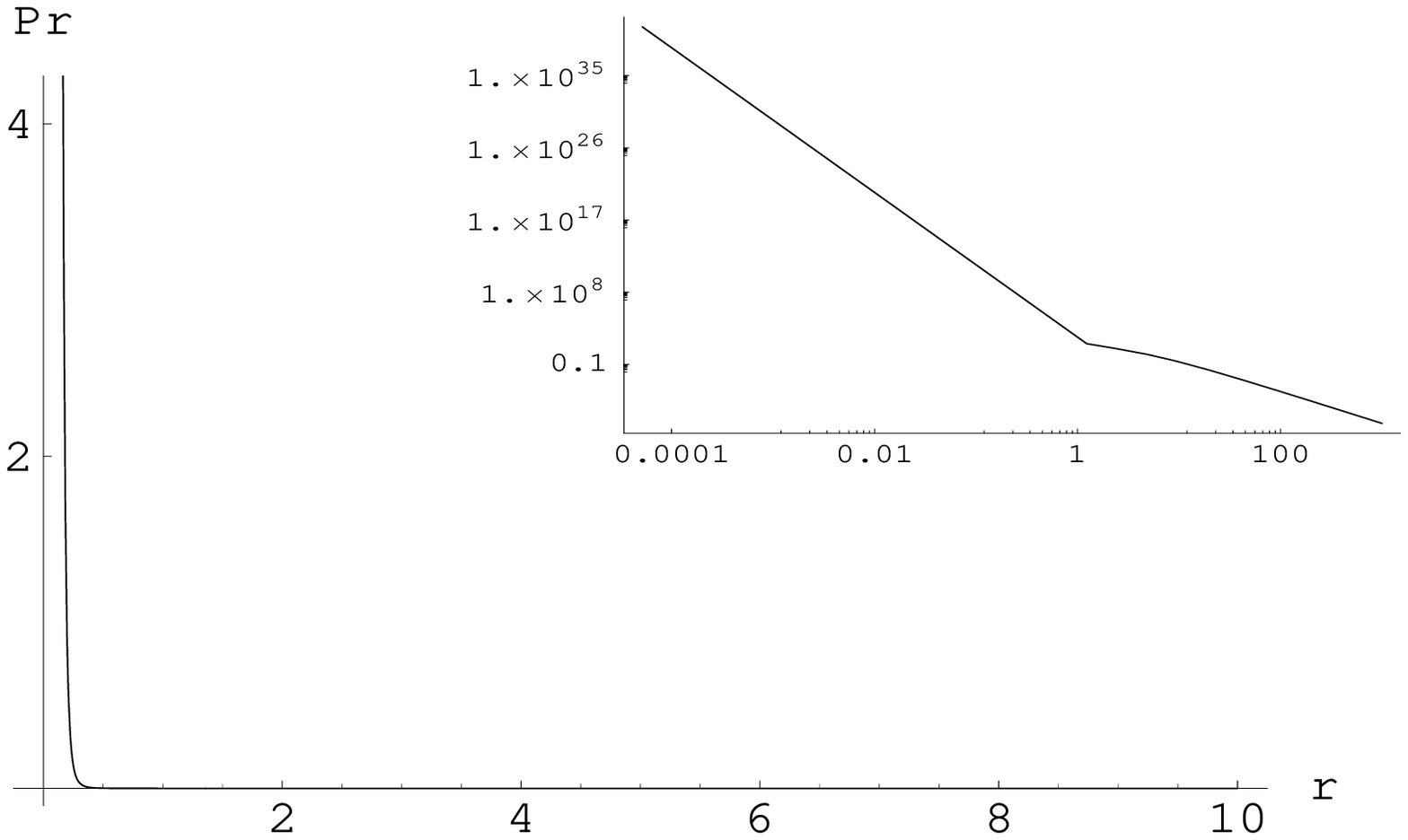}

\includegraphics[scale=0.5,angle=0]{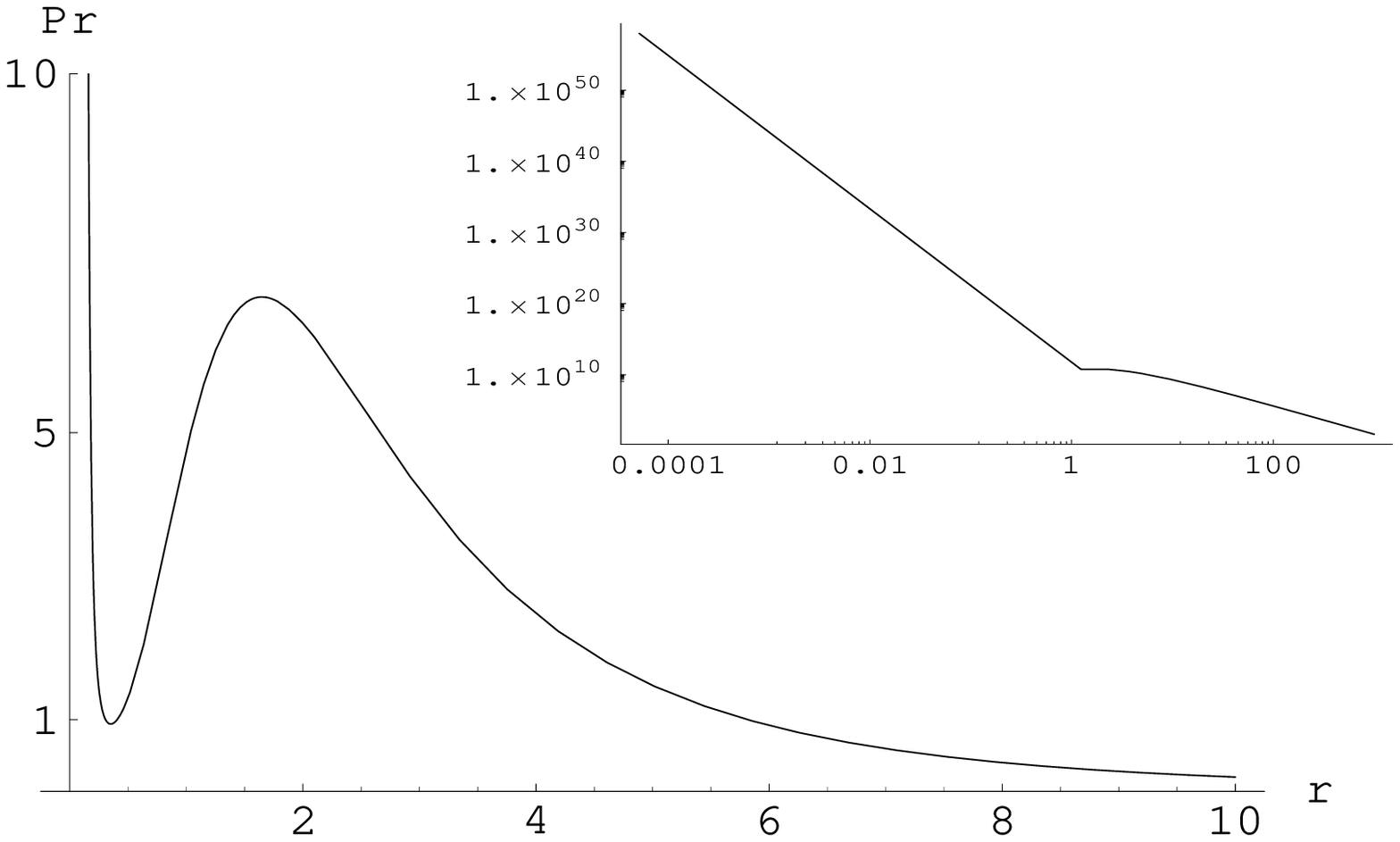}

\includegraphics[scale=0.5,angle=0]{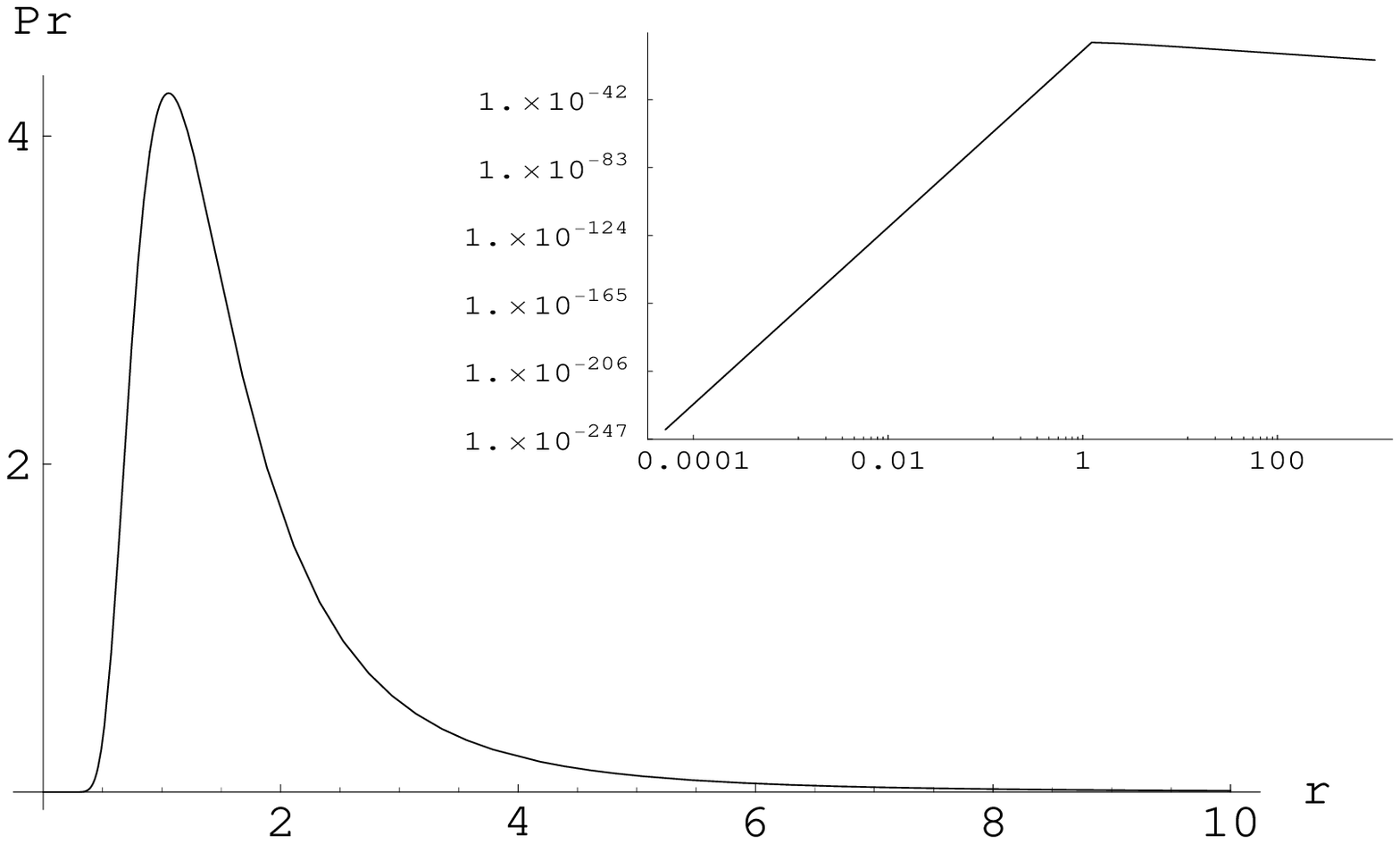}

Figure 1

\newpage
\includegraphics[scale=0.3]{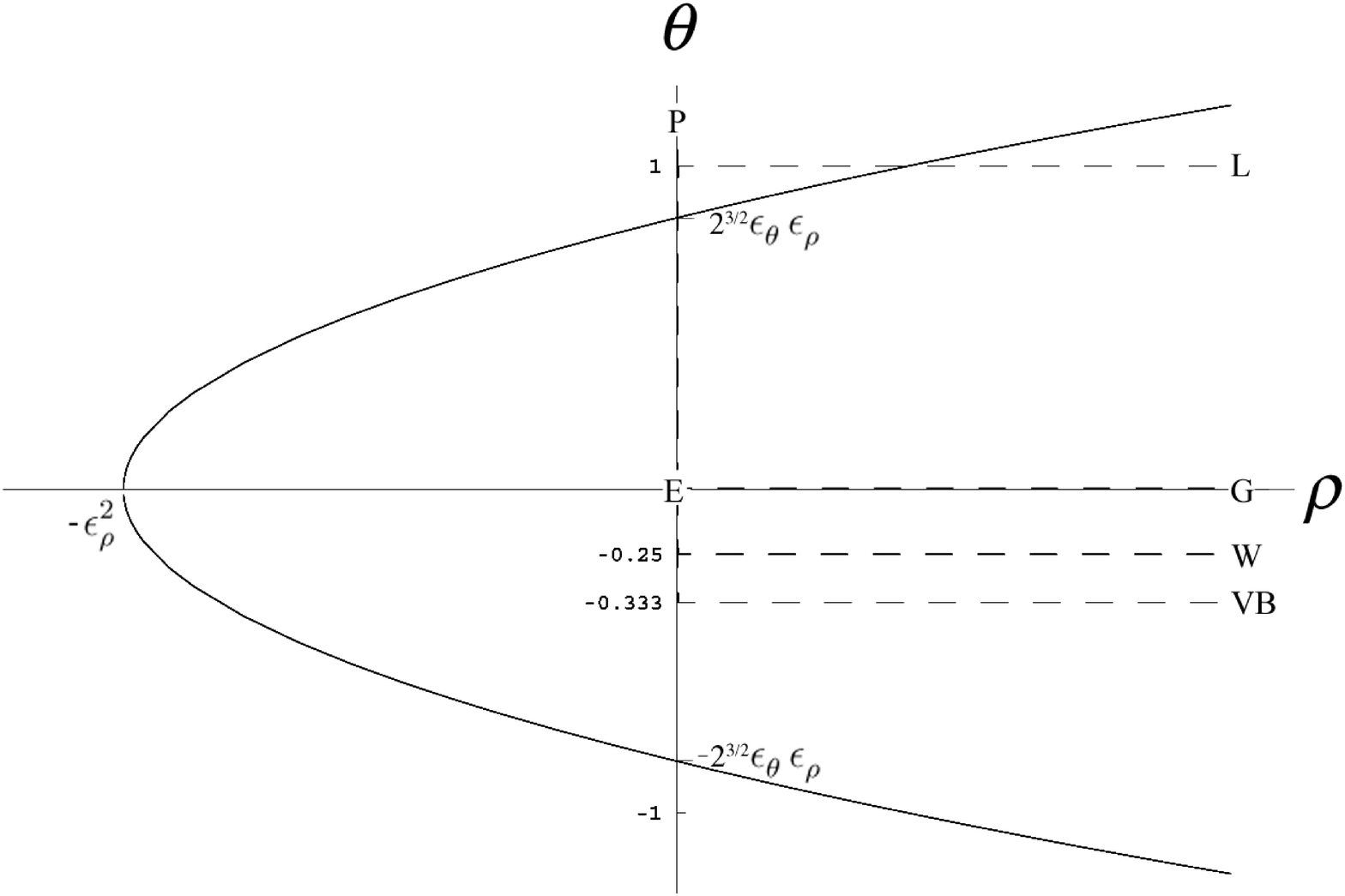}

Figure 2

\newpage
\includegraphics[scale=0.3,angle=-90]{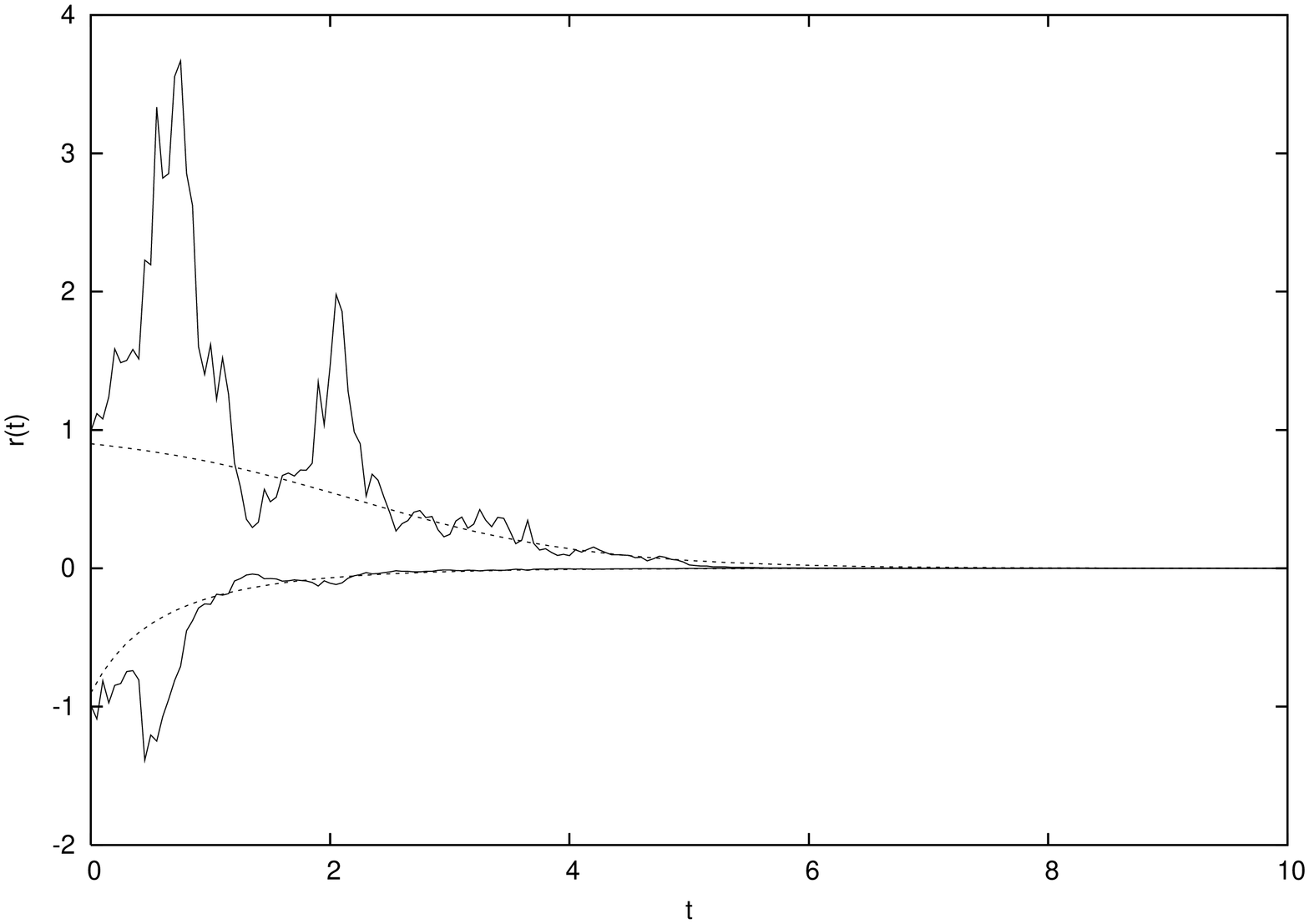}

\includegraphics[scale=0.3,angle=-90]{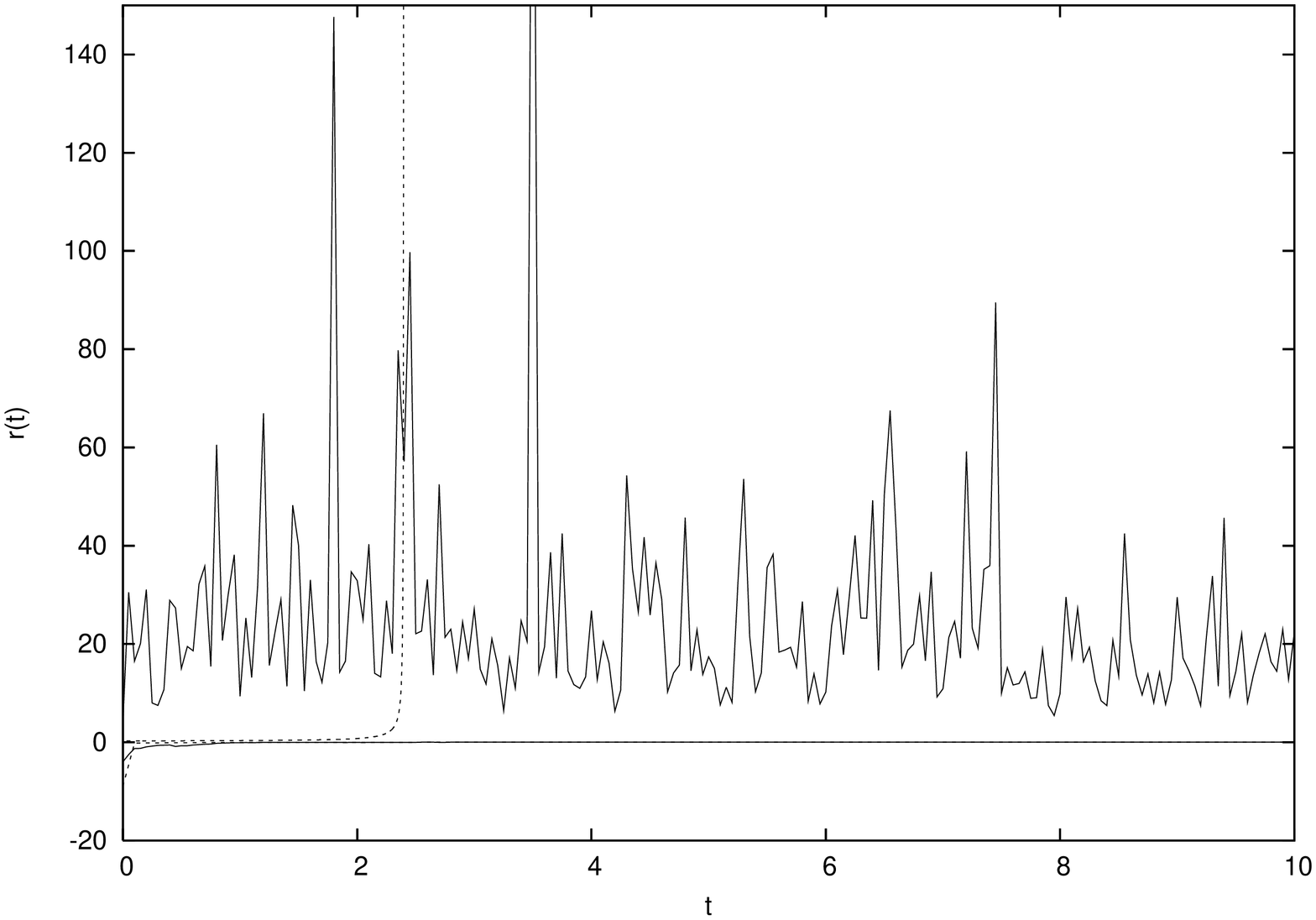}

\includegraphics[scale=0.3,angle=-90]{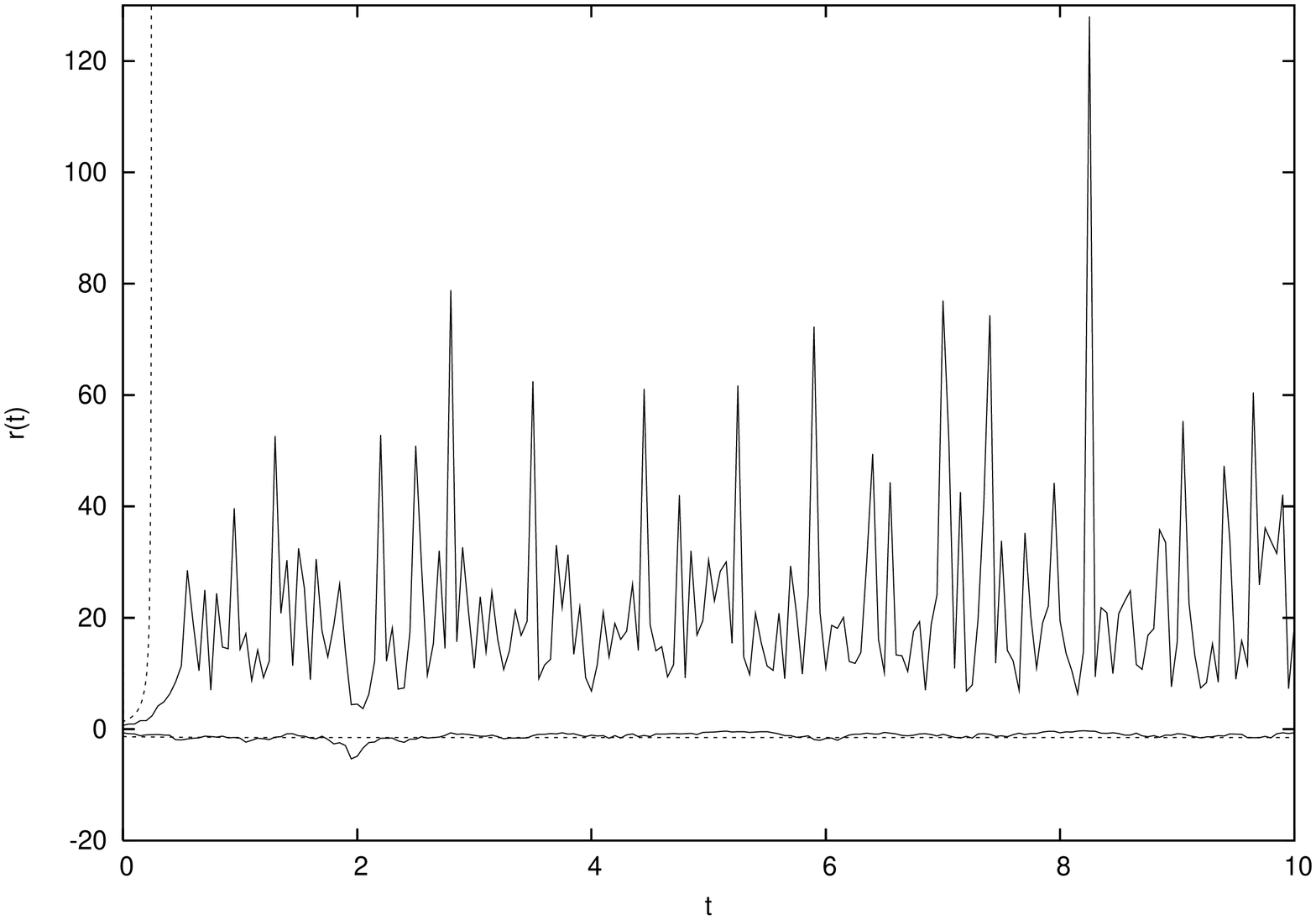}

Figure 3

\newpage
\includegraphics[scale=0.3,angle=-90]{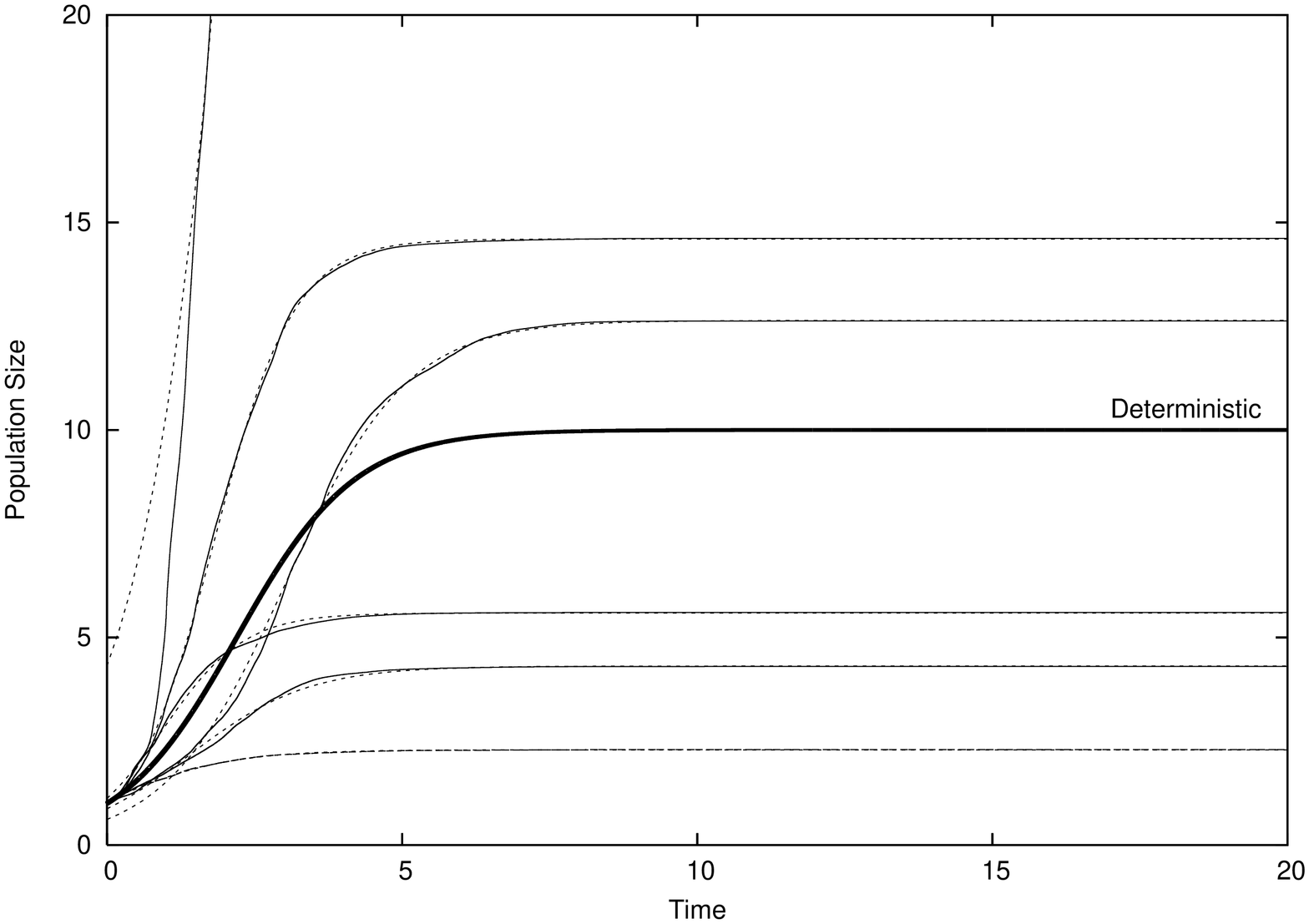}

\includegraphics[scale=0.3,angle=-90]{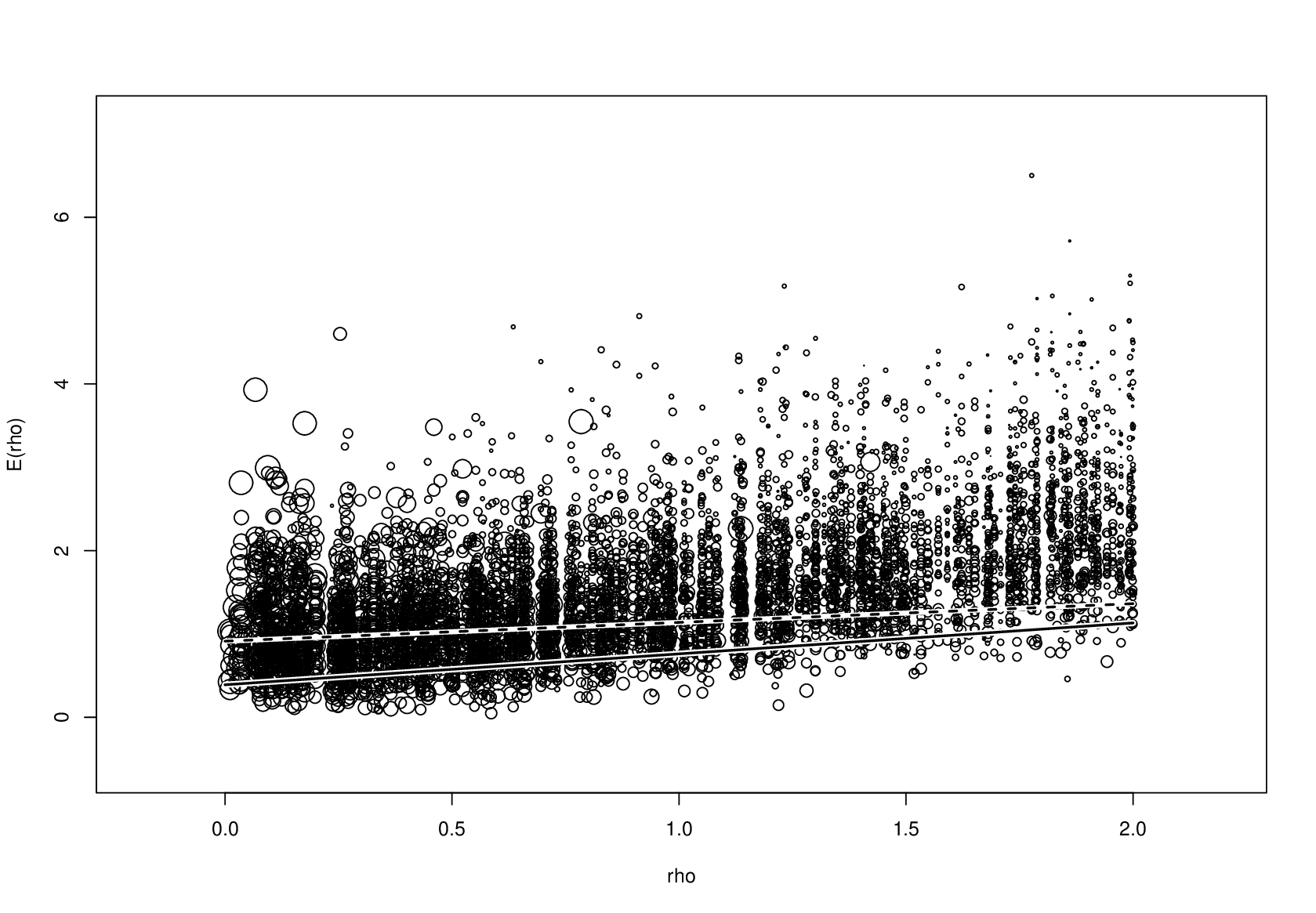}

\includegraphics[scale=0.3,angle=-90]{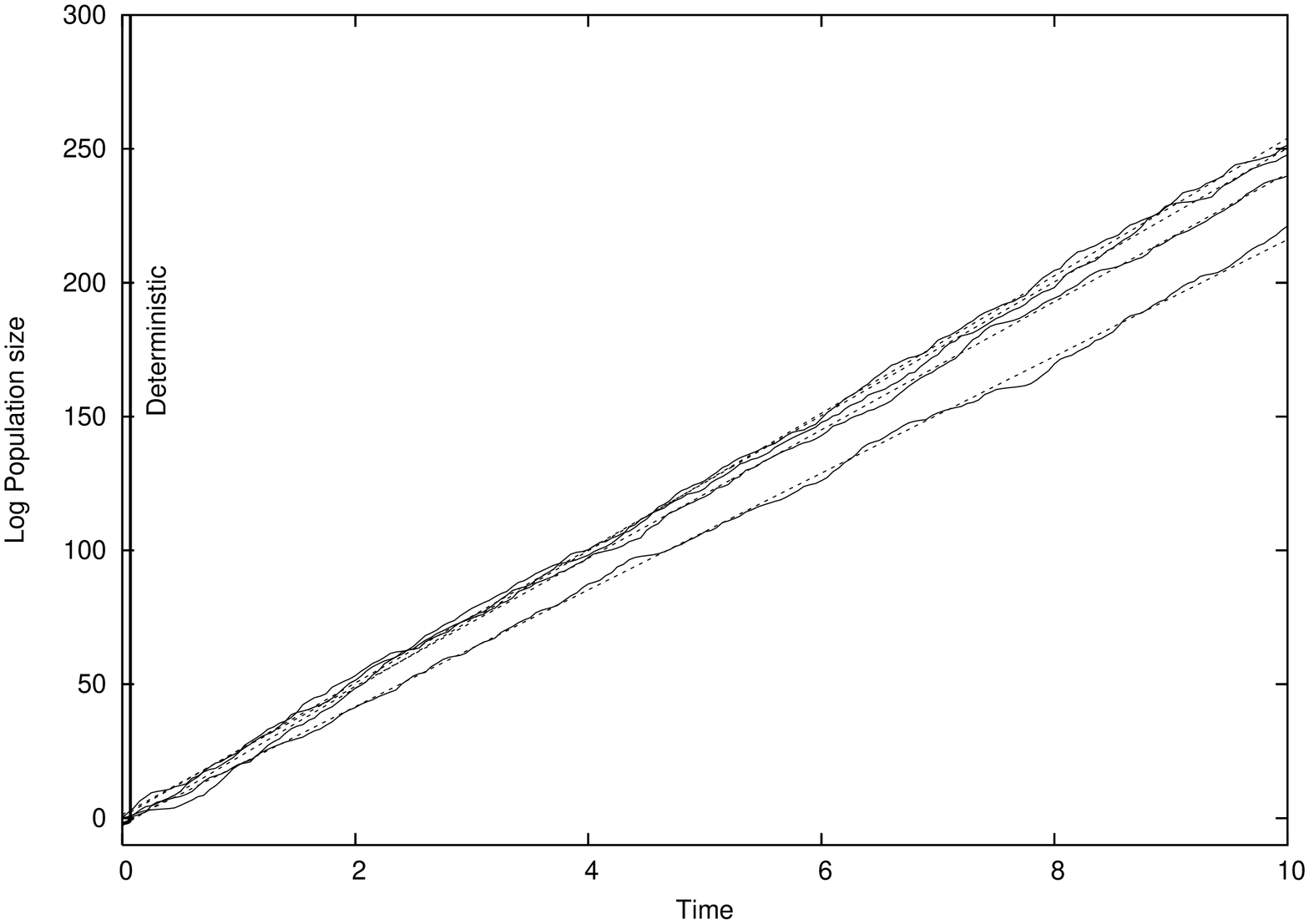}

Figure 4